\documentclass[%
reprint,
twocolumn,
superscriptaddress,
showpacs,preprintnumbers,
nofootinbib,
aps,
prd,
]{revtex4-2}

\usepackage{amsfonts}
\usepackage{bbm,bm}
\usepackage[breaklinks,urlbordercolor={1 1 1}]{hyperref}
\usepackage{graphicx}
\usepackage{amsmath}
\usepackage{amssymb}
\usepackage{mathtools}
\usepackage{hyperref}
\usepackage{graphicx}
\usepackage{color}
\usepackage{lineno}
\usepackage{xcolor}
\usepackage{comment}
\usepackage{subfigure}
\usepackage{dcolumn}
\usepackage[utf8]{inputenc} 
\usepackage{multirow}
\usepackage{cases}
\usepackage{slashed}
\usepackage{ulem}

\usepackage{orcidlink}

\newcommand{\be}{\begin{equation}}
\newcommand{\ee}{\end{equation}}
\newcommand{\ba}{\begin{array}}
\newcommand{\ea}{\end{array}}
\newcommand{\bea}{\begin{eqnarray}}
\newcommand{\eea}{\end{eqnarray}}

\usepackage{ulem,fancyvrb}
\usepackage{xcolor}

\begin{document}

\title{
Probing scalar-neutrino and scalar-dark-matter interactions with PandaX-4T
}


\def\tdli{State Key Laboratory of Dark Matter Physics, Key Laboratory for Particle Astrophysics and Cosmology (MoE), Shanghai Key Laboratory for Particle Physics and Cosmology, New Cornerstone Science Laboratory, Tsung-Dao Lee Institute, 
Shanghai Jiao Tong University, Shanghai 201210, China}
\def\sjtuphys{State Key Laboratory of Dark Matter Physics, Key Laboratory for Particle Astrophysics and Cosmology (MoE), Shanghai Key Laboratory for Particle Physics and Cosmology, School of Physics and Astronomy, Shanghai Jiao Tong University, Shanghai 200240, China}
\def\newcorner{New Cornerstone Science Laboratory, Tsung-Dao Lee Institute, Shanghai Jiao Tong University, Shanghai 201210, China}
\def\MESJTU{School of Mechanical Engineering, Shanghai Jiao Tong University, Shanghai 200240, China}
\def\SPEIT{SJTU Paris Elite Institute of Technology, Shanghai Jiao Tong University, Shanghai 200240, China}
\def\SJTUSC{Shanghai Jiao Tong University Sichuan Research Institute, Chengdu 610213, China}

\def\BUAA{School of Physics, Beihang University, Beijing 102206, China}
\def\BUAACenter{Peng Huanwu Collaborative Center for Research and Education, Beihang University, Beijing 100191, China}
\def\BUAALab{International Research Center for Nuclei and Particles in the Cosmos \& Beijing Key Laboratory of Advanced Nuclear Materials and Physics, Beihang University, Beijing 100191, China}
\def\SCNT{Southern Center for Nuclear-Science Theory (SCNT), Institute of Modern Physics, Chinese Academy of Sciences, Huizhou 516000, China}

\def\USTClab{State Key Laboratory of Particle Detection and Electronics, University of Science and Technology of China, Hefei 230026, China}
\def\USTCdep{Department of Modern Physics, University of Science and Technology of China, Hefei 230026, China}

\def\YaLongSD{Yalong River Hydropower Development Company, Ltd., 288 Shuanglin Road, Chengdu 610051, China}
\def\scKeyLab{Jinping Deep Underground Frontier Science and Dark Matter Key Laboratory of Sichuan Province, Liangshan 615000, China}

\def\pku{School of Physics, Peking University, Beijing 100871, China}
\def\CHEPpku{Center for High Energy Physics, Peking University, Beijing 100871, China}

\def\SDUdep{Research Center for Particle Science and Technology, Institute of Frontier and Interdisciplinary Science, Shandong University, Qingdao 266237, China}
\def\SDUlab{Key Laboratory of Particle Physics and Particle Irradiation of Ministry of Education, Shandong University, Qingdao 266237, China}

\def\UMD{Department of Physics, University of Maryland, College Park, Maryland 20742, USA}

\def\SYU{School of Physics, Sun Yat-Sen University, Guangzhou 510275, China}
\def\SYUSFI{Sino-French Institute of Nuclear Engineering and Technology, Sun Yat-Sen University, Zhuhai 519082, China}
\def\SYUzhuhai{School of Physics and Astronomy, Sun Yat-Sen University, Zhuhai 519082, China}
\def\SYUshenzhen{School of Science, Sun Yat-Sen University, Shenzhen 518107, China}

\def\NKU{School of Physics, Nankai University, Tianjin 300071, China}
\def\YTU{Department of Physics, Yantai University, Yantai 264005, China}
\def\FDU{Key Laboratory of Nuclear Physics and Ion-beam Application (MOE), Institute of Modern Physics, Fudan University, Shanghai 200433, China}
\def\CDUT{College of Nuclear Technology and Automation Engineering, Chengdu University of Technology, Chengdu 610059, China}

\def\AMH{On Leave from Amherst Center for Fundamental Interactions, Department of Physics, University of Massachusetts, Amherst, MA 01003, USA }
\def\CAL{Kellogg Radiation Laboratory, California Institute of Technology, Pasadena, CA 91125, USA}
\def\NCTS{Physics Division, National Center for Theoretical Sciences, National Taiwan University, Taipei 106319, Taiwan}
\def\PU{Phenikaa Institute for Advanced Study, Phenikaa University, Nguyen Trac, Duong Noi, Hanoi 100000, Vietnam}

\affiliation{\tdli}
\author{Tao Li\orcidlink{0000-0001-7225-9562}}
\affiliation{\SPEIT}
\author{Zihao Bo\orcidlink{0009-0002-0743-5368}}
\affiliation{\sjtuphys}
\author{Wei Chen\orcidlink{0009-0009-5911-7135}}
\affiliation{\sjtuphys}
\author{Xun Chen\orcidlink{0000-0001-7961-7908}}
\affiliation{\tdli}\affiliation{\SJTUSC}
\affiliation{\scKeyLab}
\author{Yunhua Chen}\affiliation{\YaLongSD}
\affiliation{\scKeyLab}
\author{Chen Cheng\orcidlink{0000-0003-0164-7538}}
\affiliation{\BUAA}
\author{Xiangyi Cui}
\affiliation{\tdli}
\author{Manna Deng}
\affiliation{\SYUSFI}
\author{Yingjie Fan}
\affiliation{\YTU}
\author{Deqing Fang}
\affiliation{\FDU}
% ...
\collaboration{PandaX Collaboration}
\author{Yihong Zhong\orcidlink{0009-0006-6414-1920}}
\email[Corresponding author: ]{yihong-z@sjtu.edu.cn}
\affiliation{\tdli}
\affiliation{\sjtuphys} 
\author{Van Que Tran\orcidlink{0000-0003-4643-4050}}
\email[Corresponding author: ]{vqtran@phys.ncts.ntu.edu.tw}
\affiliation{\tdli}
\affiliation{\NCTS}
\affiliation{\PU}
\author{Michael J.~Ramsey-Musolf\orcidlink{0000-0001-8110-2479}}
\email[Corresponding author: ]{mjrm@sjtu.edu.cn}
\affiliation{\tdli}
\affiliation{\sjtuphys}
\affiliation{\AMH}
\affiliation{\CAL}  
\noaffiliation

\begin{abstract}

Scalar-mediated interactions may exist among neutrinos, dark matter particles, or between the two. Double $\beta$-decay experiments provide a powerful tool to probe such exotic interactions. Using $^{136}$Xe double $\beta$-decay data from PandaX-4T, we perform the first direct spectral search in the energy range of 20 to 2800~keV, setting the most stringent limits to date on scalar-mediated neutrino self-interactions for mediator masses below 2~MeV$/c^2$. These results place significant constraints on models invoking such interactions to alleviate the Hubble Tension. Assuming the same scalar also mediates dark matter self-interactions, constraints on the dark matter-scalar interactions can be placed in conjunction with cosmological constraints.
\end{abstract}

\maketitle

\renewcommand{\linenumbersep}{3pt}

The standard cosmological model, $\Lambda$CDM, has been remarkably successful in describing the evolution and structure of the universe~\cite{Trujillo-Gomez:2010jbn,Springel:2006vs,Bahcall:1999xn}. However, recent observational tensions challenge its completeness. One prominent example is the Hubble tension, a persistent discrepancy between measurements of the Hubble constant derived from early-universe observations and those obtained from late-universe data \cite{Riess:2018kzi, Planck:2018vyg, Riess:2019cxk, Shanks:2018rka, Riess:2016jrr}.
Furthermore, recent James Webb Space Telescope observations disfavor dominant late-time systematics \cite{Riess:2024ohe,Anand:2024nim,Li:2024yoe,Freedman:2023jcz}, strengthening the case for new physics beyond $\Lambda$CDM in the early and/or late Universe \cite{Vagnozzi:2023nrq}.
Among the proposed solutions, neutrino self-interactions ($\nu$SI) mediated by a light boson have attracted considerable attention~\cite{Cyr-Racine:2013jua,Kreisch:2019yzn,Lancaster:2017ksf,Oldengott:2017fhy}.
Notably, the presence of a new mediator boson could also produce a distinct signature in the nuclear double $\beta$-decay process, manifesting as anomalous missing energy beyond the expectations of standard decay processes, as first highlighted in Ref.~\cite{Georgi:1981pg}. The search for such novel neutrino interactions is compelling in its own right. For instance, a specific class of scalar boson, known as the Majoron, emerges in models where lepton number symmetry is spontaneously broken—often in connection with neutrino mass generation mechanisms, such as the seesaw models~\cite{Yanagida:1979as, Glashow:1979nm, Mohapatra:1979ia, Schechter:1980gr}.

In addition to the Hubble tension, the $\Lambda$CDM model faces challenges on smaller cosmological scales, including discrepancies between observations and the modeling of dwarf galaxies, galaxies, and galaxy clusters—collectively referred to as the ``small-scale problems". A promising solution to these issues involves introducing self-interactions among dark matter particles (SIDM), mediated by a light boson\cite{Feng:2009hw,Tulin:2013teo,Ko:2014nha,Boddy:2014yra,Tulin:2017ara}. If this mediator is the same that mediates neutrino self-interactions, it naturally gives rise to neutrino-dark matter ($\nu$-DM) interactions. By combining cosmological constraints on $\nu$-DM interactions with $\nu$SI signatures in double $\beta$-decay data, one can place constraints in the SIDM parameter space.

In this Letter, we report the most sensitive search to date for the $\nu$SI scalar mediator ($\phi$) with a mass below 2~MeV/$c^2$, based on $^{136}$Xe double $\beta$-decay data with an exposure of 37.8~kg·yr from the PandaX-4T experiment. 
Furthermore, we place constraints on SIDM by combining cosmological bounds on $\nu$-DM interactions for a DM candidate with mass below 100 GeV/$c^{2}$, under the assumption that $\phi$ is also the mediator for SIDM.

At low energies, $\nu$SI are often described by an effective four-fermion operator, ${\cal L}_{\text{eff}} \supset G_{\rm eff}(\bar{\nu}\nu)(\bar{\nu}\nu)$, 
obtained by integrating out the mediator exchanged in ultraviolet (UV)-complete theories.
This contact-interaction approximation is valid only when the typical momentum transfer is much smaller than the mediator mass. 
In the early universe, particularly during epochs relevant to the Hubble tension ($T \sim 1\text{ eV}$–$1\text{ keV}$), the neutrino scattering energy can approach the mediator mass~\cite{Blinov:2019gcj}. 
This causes the effective description to break down, requiring the mediator to be treated explicitly to ensure an accurate description of the dynamics.

Furthermore, realizing SIDM phenomenology requires a light mediator with mass below $\mathcal{O}(10~\mathrm{MeV}/c^{2})$, capable of generating the long-range forces needed to reproduce the observed small-scale dynamics~\cite{Tulin:2017ara}.
Therefore, to consistently account for both $\nu$SI relevant to the Hubble tension problem and SIDM to deal with smaller scale issues within a unified framework, one must introduce a light mediator that couples to both the neutrino and dark sectors.
Such a particle can naturally arise in various UV completions~\cite{Batell:2017cmf,Olivares-DelCampo:2017feq,Blennow:2019fhy}, linking the neutrino and DM interactions through a common mediator portal.

Both vector and scalar mediators can in principle produce the required $\nu$SI and SIDM couplings.
However, the scalar case offers the simplest Lorentz-invariant structure, and constraints from Big Bang nucleosynthesis (BBN) are generally more stringent for vector or complex scalar mediators due to their additional degrees of freedom.
We therefore focus on a minimal scenario involving a real scalar field $\phi$ with mass $m_\phi$, mediating interactions between neutrinos and dark matter.
We adopt an effective description in which neutrinos are Majorana and the dark matter $\chi$ is a standard-model-singlet Dirac fermion\footnote{We assume that the Majorana nature of neutrinos and the Dirac nature of the DM fermion originate from an underlying UV-complete framework. For instance, Majorana neutrino masses may arise from lepton-number-violating dynamics such as seesaw mechanisms~\cite{Minkowski:1977sc, Mohapatra:1979ia}, while the Dirac nature of $\chi$ can be ensured by a conserved dark sector symmetry, as commonly realized in extended gauge or scotogenic-type models~\cite{Ma:2006km}.}.
The relevant Lagrangian is
\be
    \label{eq:Lagrangian in letter}
    {\cal L} = g_{\nu \phi}   \overline{\nu}P_R\nu^c \phi +  g_{\chi \phi} \overline{\chi} \chi \phi + {\rm h.c.,}
\ee
where $g_{\nu \phi}$ and $g_{\chi \phi}$ are the corresponding coupling constants. We note that the case of bosonic dark matter can be treated analogously. 
The resulting Yukawa potential between dark matter particles from~Eq.~(\ref{eq:Lagrangian in letter}) is purely attractive, enabling self-interacting dark matter with a velocity-dependent cross section consistent with astrophysical observations~\cite{Buckley:2009in,Feng:2009hw,Loeb:2010gj}.

\begin{figure}
    \centering
    \includegraphics[width=0.7\columnwidth]{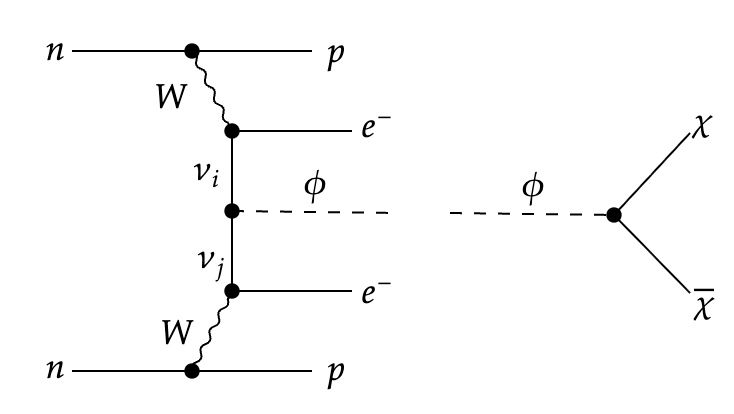}
    \caption{
    The Feynman diagram for the double $\beta$-decay process accompanied by a scalar emission, which can be detected in PandaX-4T experiment (left). The same scalar may also mediate the interactions between dark matter particles (right).}
    \label{fig:Feynman_diagram}
\end{figure}

In this model, the neutrinoless double $\beta$-decay process can proceed through real $\phi$ emission, as illustrated in the the left diagram of Fig.~\ref{fig:Feynman_diagram}, leading to a characteristic missing energy. 
The emitted $\phi$ may subsequently decay into a pair of sufficiently light dark matter particles or neutrinos, neither of which are detectable.

The decay rate for the double $\beta$-process accompanied by $\phi$ emission is given by:
\be
    \label{eq:decayrate}
    \Gamma_{\beta\beta\phi} = \left({g_{\nu\phi}} \right)^2 \vert M_{\nu\phi} \vert^2 {\cal G}_{0\nu,\phi},
\ee
where $M_{\nu\phi}$ is the nuclear matrix element (NME) (see below) and ${\cal G}_{0\nu,\phi}$ is the modified phase space factor ruling such decay. The detailed calculation and formule for them can also be found in Supplemental Material~\cite{SupplementalMaterial}.

The NME for double $\beta$-decay with massless Majoron emission, via the left diagram in Fig.~\ref{fig:Feynman_diagram}, has been extensively studied in Refs.~\cite{Doi:1985dx,Doi:1987rx,Mohapatra:1988fk}.
In our case, we generalize this framework to include a massive scalar $\phi$, whose mass lies below the typical double $\beta$–decay energy scale, i.e., $m_{\phi} \lesssim \mathcal{O}(\text{MeV})$. Following Ref.~\cite{Doi:1985dx}, the $S$ matrix element can be decomposed into two parts using the leptonic Dirac algebra as shown in Supplemental Material~\cite{SupplementalMaterial}.
The NME can then be written as:
\begin{equation}
    \begin{aligned}
        \label{eq:NME}
        |M_{\nu\phi}| = &
         \frac{2\pi R_0}{g_A^2} \int d^{3}x \int d^{3}y 
         \sum_{N}  {\cal K}^{(1)}(\vec{x}, \vec{y}) \\
         & \; \langle F \vert \mathcal{J}^{\rho\dagger}_{L}(\vec{x}) \vert N \rangle \langle N \vert \mathcal{J}^{\dagger}_{\rho,L}(\vec{y}) \vert I \rangle .
    \end{aligned}
\end{equation}
Here, $g_A \simeq 1.27$ is the axial-vector coupling of the hadronic current~\cite{ParticleDataGroup:2014cgo}, and the nuclear radius $R_0$ is included to ensure the correct dimensionality. 
As analyzed in Supplemental Material~\cite{SupplementalMaterial}, we follow the same setup in \cite{EXO-200:2014vam, KamLAND-Zen:2012uen} and take
\begin{equation}
|M_{\nu\phi}| \simeq |M_{0\nu}| = 3.11,
\end{equation}
using the numerical NME value from Ref.~\cite{Mustonen:2013zu,Simkovic:2009pp} with quasiparticle random phase approximation method. 
Values from other nuclear calculations range from 2.59 to 3.11~\cite{Simkovic:2009pp}, resulting in an uncertainty of approximately 20\% in the NME calculation.

\begin{figure}[htb]
   \centering
   \includegraphics[width=1.\columnwidth]{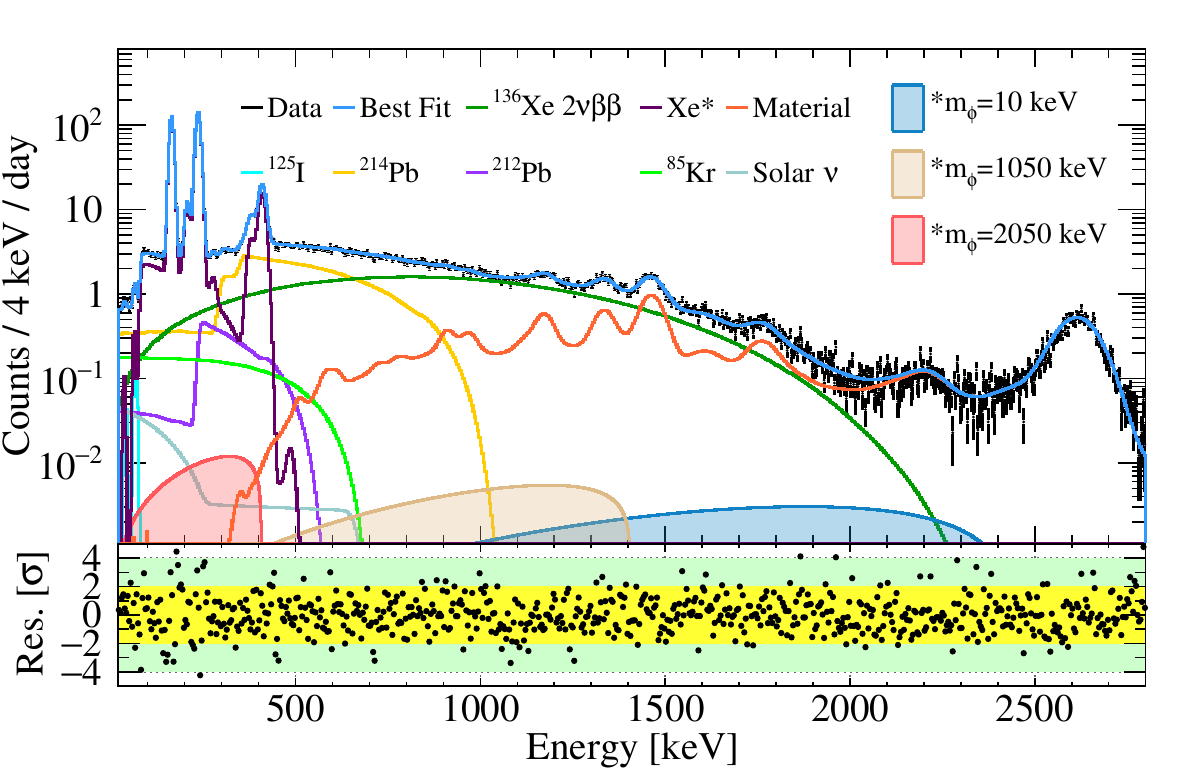}
    \caption{
    The background-only fit to the combined SS data from Run0 and Run1, spanning from 20 to 2800 keV with a bin size of 4 keV. 
    The contributions from xenon isotopes are included in the Xe$^*$ term. 
    The lower panel displays the residuals, with the corresponding $\pm 2 \sigma$ and $\pm 4 \sigma$ bands. 
    Additionally, the hypothetical spectra for mediator masses $m_{\phi} = 10$ keV, $m_{\phi} = 1050$ keV and $m_{\phi} = 2050$ keV, with the corresponding values of $g_{\nu\phi}$ at the PandaX-4T upper limits, $1.4 \times 10^{-5}$, $3.3 \times 10^{-5}$, and $2.4\times 10^{-4}$, are also shown.
     }
    \label{fig:fit_result}
\end{figure}

The dominant contribution to the spectrum comes from the standard two-neutrino double $\beta$-decay process, but the emission of a massive scalar $\phi$ exhibits distinctive features (see in Supplemental Material~\cite{SupplementalMaterial} for the analytical expression of $\mathcal{G}_{0\nu,\phi}$, which has the upper limit of $3.1\times 10^{-47}$ GeV when the mediator mass approaches zero). As $m_{\phi}$ increases, the total energy spectrum of the two $\beta$s shifts to lower energies, as illustrated in Fig.~\ref{fig:fit_result}, clearly distinguishing it from the standard double $\beta$-decay spectrum. 
On the experimental side, the PandaX-4T experiment reports a measurement of the double $\beta$-decay half-life among the most precise to date~\cite{PandaX:2025yly}.
We shall therefore use the dataset of double $\beta$-decay events from $^{136}$Xe, collected by PandaX-4T, to conduct a highly sensitive search.

The PandaX-4T experiment is located at the China Jinping Underground Laboratory Phase II (CJPL-II) in Sichuan, China.
It utilizes a dual-phase time projection chamber (TPC) filled with natural xenon (with a measured $^{136}$Xe abundance of 8.6\%) to search for dark matter (DM) and neutrino-related rare events.
The TPC has a cylindrical structure with a diameter and height of 118.5 cm, surrounded by 24 reflective panels, and contains 3.7 tonnes of natural xenon.
A uniform drift field is established between the cathode at the bottom and the gate electrode at the top, with the anode positioned 10 mm above the gate.
Photon signals are detected by two arrays of photomultiplier tubes: 199 three-inch Hamamatsu R11410-23 at the bottom and 169 at the top.
Further details on the detector design and operational parameters can be found in Refs.~\cite{PandaX-4T:2021bab, PandaX:2024qfu}.
In this work, we utilize the commissioning dataset (Run 0) and the first scientific dataset (Run 1), collected between November 2020 and May 2022, to probe the neutrino-scalar interaction.

The detector measures the deposited energy and the three-dimensional event position using the scintillation signal (S1) and the electroluminescence signal (S2), the latter of which scales with the number of ionized electrons.
Data processing, including position and energy reconstruction as well as event selection, follows the procedures outlined in Refs.~\cite{PandaX:2024sds, PandaX:2024fed}.
The horizontal position is reconstructed using a maximum likelihood estimation with the charge pattern of S2 in the top PMT array and the photon acceptance functions derived from optical Monte Carlo simulations.
The detector's energy response, including the energy scale and resolution, is parameterized using two-parameter and three-parameter functions of the visible energy, respectively.
The calibrated parameters, along with their associated uncertainties, are incorporated into the likelihood function for spectrum fitting.
Single-site (SS) events, defined as those with a single reconstructed S2 cluster, are selected within an energy region of interest (ROI) of 20–2800~keV.
The fiducial volume, defined as the innermost cleanest part of the detector in Ref.~\cite{PandaX:2024sds}, corresponds to an exposure of 37.8~kg·yr of $\mathrm{^{136}Xe}$.

\begin{figure}[htb]
   \centering
   \includegraphics[width=1.\columnwidth]{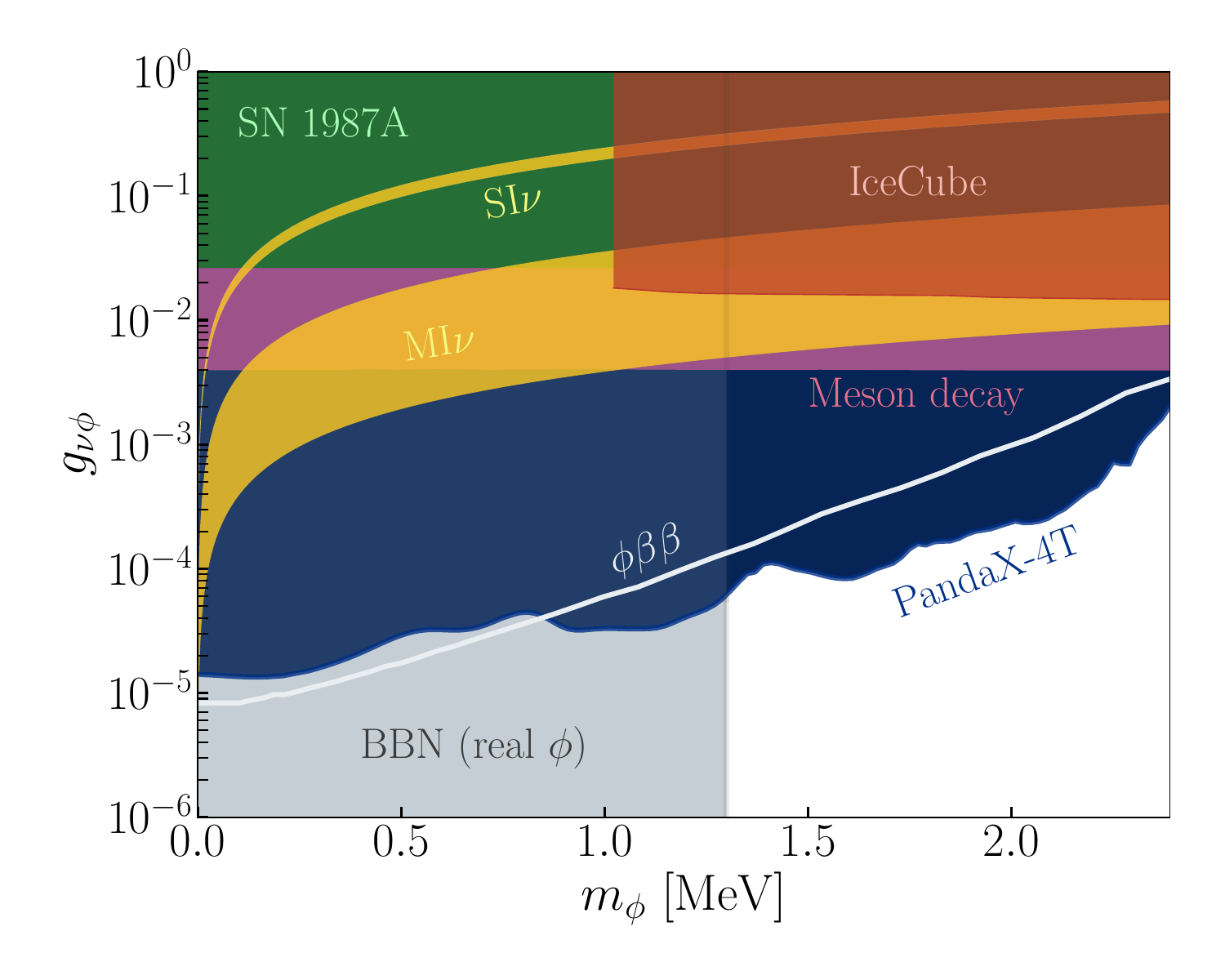}
    \caption{ Bounds on $g_{\nu\phi}$ with varing mass of $\phi$.
    The blue region presents the direct probing upper limit on $g_{\nu\phi}$ from PandaX-4T experiment.
    The white solid line stands for the re-interpreted constraints from the massless Majoron limits using the two-neutrino double $\beta$-decay data of $^{48}$Ca, $^{100}$Mo, $^{136}$Xe and $^{150}$Nd~\cite{Brune:2018sab}.
    The excluded region from BBN nucleosynthesis constraints with a preferred value of the baryon density for real scalar mediators~\cite{Blinov:2019gcj} is shown in grey vertical band.
    The light yellow bands presents the preferred parameter space for the strong interaction (SI$\nu$) and the moderate interaction (MI$\nu$) case~\cite{Anand:2024nim}, respectively~\cite{Kreisch:2019yzn}.
    We also provide some references for relevant constraints from Meson decay~\cite{Berryman:2018ogk} (dark blue), SN 1987 A ~\cite{Kolb:1987qy,Shalgar:2019rqe} (green) and IceCube ~\cite{Hyde:2023eph,IceCube:2022der} (orange band).
    }
    \label{fig:constraints_on_gnuchi}
\end{figure}

The background modeling and constraints within the ROI are consistent with those detailed in Refs.\cite{PandaX:2024sds, PandaX:2024fed}, encompassing contributions from the outer structure, solar neutrinos, and liquid xenon.
The background from the outer structure, which includes the detector materials and the stainless steel platform, is estimated using data from the high-purity germanium counting station\cite{PandaX:2022kwg} and multi-site event analysis~\cite{PandaX-4T:2025jel}, respectively.
A time evolution analysis is employed to constrain Gaussian mono-energetic peaks from short-lived xenon isotopes ($\mathrm{^{127}Xe}$, $\mathrm{^{129m}Xe}$, and $\mathrm{^{131m}Xe}$)\cite{PandaX:2024sds}.
The level of $\mathrm{^{85}Kr}$ is determined via $\beta$-$\gamma$ cascades through the isomeric state of $\mathrm{^{85m}Rb}$\cite{PandaX:2024qfu}.
The levels of $\mathrm{^{124}Xe}$ and $\mathrm{^{125}I}$ are taken from previously reported PandaX-4T measurements~\cite{PandaX-4T:2024fls}.
The remaining xenon isotopes ($\mathrm{^{125}Xe}$ and $\mathrm{^{133}Xe}$), as well as $\mathrm{^{212}Pb}$ and $\mathrm{^{214}Pb}$, are treated as free parameters in the likelihood fit. The primary contribution to the fit comes from the standard double $\beta$-decay of $\mathrm{^{136}Xe}$, 
with its rate determined based on the half-life measurement by PandaX-4T~\cite{PandaX:2022kwg}
with its rate left floating in the likelihood fit.
Additional minor physical backgrounds include solar neutrino-electron scattering events ($pp$ and $\mathrm{^{7}Be}$), with their spectra and associated uncertainties adopted from Refs.~\cite{Chen:2016eab, BOREXINO:2014pcl}.

The binned-likelihood spectral fit is performed for different mediator masses, as well as for the massless scenario, with the signal amplitude allowed to float.
Under the on-shell mediator scenario, and accounting for the detector energy resolution and efficiency, the mediator mass $m_{\phi}$ is scanned from 10 keV to 2390 keV in steps of 20 keV.
Gaussian penalty terms are applied to constrain the nuisance parameters, which account for uncertainties in the detection efficiency, event selection, and background modeling.
A convolution method is employed to propagate the uncertainties of the energy response into the energy spectrum, ensuring that all relevant uncertainties are fully incorporated in the likelihood fit.
Additional details on the statistical approach can be found in Ref.~\cite{PandaX:2024sds}.
Figure~\ref{fig:fit_result} shows the background-only fit to the SS energy spectrum, with contributions from other background components consistent with their expectations. No significant excess above the expected background was observed.

Therefore, 90\% confidence level upper limits on the event rate were determined and subsequently converted into constraints on $g_{\nu\phi}$, as shown in Fig.~\ref{fig:constraints_on_gnuchi}. For comparison, we also overlay the results from a  phenomenological analysis in Ref.~\cite{Brune:2018sab} (white curve), a re-interpretation of the world limits on massless Majoron using two-neutrino double $\beta$-decay data but ignoring the energy dependence of background in those experiments. 
In contrast, our limit is obtained through a complete spectral fit on the PandaX-4T data, resulting into a tightest bound for a mediator mass above 0.8~MeV.
Consistent with the findings in Refs.~\cite{Kreisch:2019yzn,Blinov:2019gcj,Deppisch:2020sqh}, our results challenge neutrino self-interaction models proposed to resolve the Hubble tension. Such models typically require neutrino–mediator couplings of $g_{\nu\phi} \sim (0.4 - 4)\times10^{-2}$ (“moderately interacting” (MI$\nu$)) and $g_{\nu\phi} \sim (2 - 3)\times10^{-1}$ (“strongly interacting” (SI$\nu$)) for a mediator mass of $m_{\phi} \simeq 1$ MeV\cite{Kreisch:2019yzn}. 
Within the context of the light scalar exchange paradigm discussed here, the tension between small scale structure observations, the Hubble constant, and the PandaX constraints persists and could point to additional new physics beyond the $\Lambda$CDM. For example, alternative explanations such as early dark energy, modified recombination histories, and extended dark-sector dynamics have been proposed to alleviate the Hubble tension \cite{Verde:2019ivm, Poulin:2018cxd, DiValentino:2021izs, Chluba:2019nxa}.

Also overlaid in Fig.~\ref{fig:constraints_on_gnuchi} is the Big Bang Nucleosynthesis (BBN) bound on a real scalar mediator. This bound comes from limits on the effective neutrino flavor number $N_{\rm eff}$~\cite{Blinov:2019gcj}, which may be relaxed if the energy density of light particles is diluted relative to the SM radiation, for example, through the late decay of a heavy particle into the SM bath after the light particles freeze out~\cite{Bleau:2023fsj}.

\begin{figure}[t]
   \centering
   \includegraphics[width=1.\columnwidth]{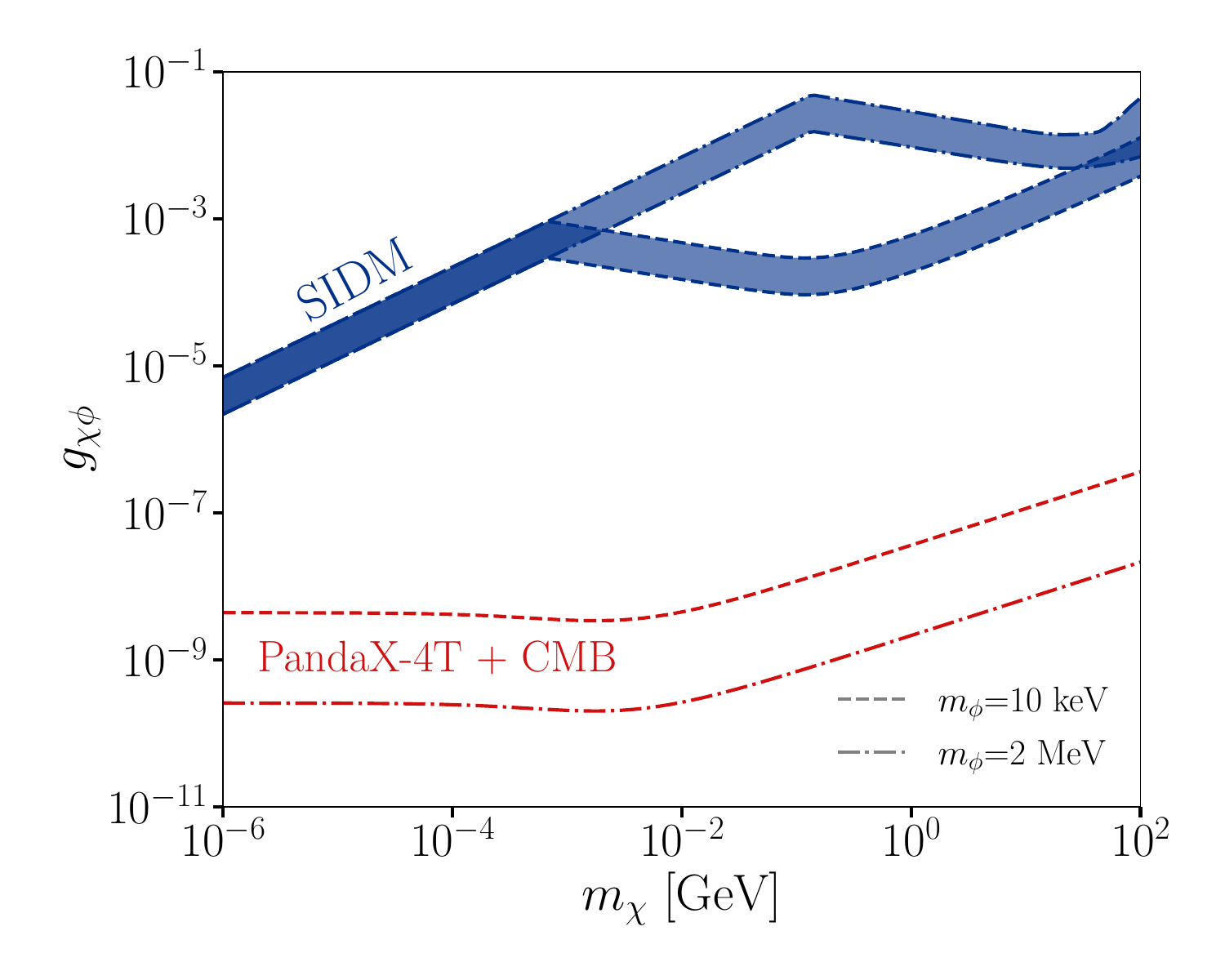}
    \caption{
    Constraints on the coupling $g_{\chi\phi}$ as a function of DM mass. The blue bands (dash-dotted and dashed) show the ranges favored by SIDM, corresponding to dwarf-scale observations with $\sigma_T/m_{\chi} = 0.1$–$10~\mathrm{cm^2/g}$ for mediator masses of $m_{\phi} = 10~\mathrm{keV}$ (dashed) and $m_{\phi} = 2~\mathrm{MeV}$ (dash-dotted).
    The dashed-dotted/dashed red lines represent the upper limit of $g_{\chi\phi}$ using CMB constraints on the DM-neutrino interaction (Eq.~(\ref{eq:sig_nu_chi})) with the same choose of the mediator masses: $m_{\phi} = 10~\mathrm{keV}$ (dashed) and $m_{\phi} = 2~\mathrm{MeV}$ (dash-dotted), assuming that $g_{\nu\phi}$ takes the value of the current PandaX-4T upper limit.
    }
    \label{fig:CMB_PandaX}
\end{figure}

Assuming that dark matter (DM) self-interaction ($\chi\chi$) is also mediated by the exchange of the same scalar $\phi$, astrophysical observations can provide strong constraints on $g_{\chi\phi}$, the coupling constant between DM and $\phi$.
It has been shown that if $\sigma_T / m_\chi$, the DM self-interaction cross section per unit DM mass, lies within the range of $\sim 0.1 - 10~\mathrm{cm^2/g}$, it can flatten the central density profiles of dwarf galaxies, potentially resolving the small-scale structure problem~\cite{Spergel:1999mh, Rocha:2012jg, Vogelsberger:2012ku}, while remaining consistent with observational bounds on Milky Way and galaxy cluster scales.
Using the numerical methods described in Ref.~\cite{Colquhoun:2020adl}, we derived the allowed parameter space for $g_{\chi\phi}$ across different mediator masses in SIDM senario, considering dwarf galaxy-scale self-interaction cross sections and the corresponding DM scattering velocities, $v_\mathrm{dw} = 10~\mathrm{km/s}$. 
The resulting constraints on the coupling as a function of dark matter mass are shown as the blue bands in Fig.~\ref{fig:CMB_PandaX}, for two representative mediator masses (10 keV and 2 MeV) probed by PandaX.

The $\nu\phi$ and $\chi\phi$ couplings inevitably introduce interactions between the DM and neutrinos. Such interactions, if present, would leave observable imprints in cosmological observations. 
Specifically, if DM scatters off relativistic neutrinos during the radiation-dominated era, it suppresses primordial density fluctuations, erases small-scale structures, and produces observable signatures in the cosmic microwave background (CMB) and the matter power spectrum \cite{Cline:2022qld, Mangano:2006mp, Wilkinson:2014ksa}.
In this work, we adopt the limit on the energy-squared dependent scattering cross section and translate it to a fixed neutrino energy $E_\nu = Q/2 \cong 1.229$ MeV, obtaining
\be
    \label{eq:sig_nu_chi}
    \sigma_{\nu\chi}^{\rm el} \lesssim 10^{-30} (m_{\chi}/{\rm GeV}) {\rm cm}^2
\ee
for non-zero neutrino masses~\cite{Mosbech:2020ahp, Paul:2021ewd}
(the limit becomes stronger by about three orders of magnitude for massless neutrinos~\cite{Wilkinson:2014ksa}). On the other hand, 
a value of $g_{\nu\phi}$ can be combined with Eq.~(\ref{eq:sig_nu_chi}) to derive corresponding upper bounds on $g_{\chi\phi}$ (see Supplemental Material~\cite{SupplementalMaterial}). 
For illustration, assuming $g_{\nu\phi}$ takes the current upper limit value in Fig.~\ref{fig:constraints_on_gnuchi}, 
the red dashed curves in Fig.~\ref{fig:CMB_PandaX} show the corresponding upper bounds on $g_{\chi\phi}$ for two representative mediator masses (10 keV and 2 MeV) probed by PandaX. 
If $g_{\nu\phi}$ saturates its current experimental upper limit, the allowed region becomes largely incompatible with the SIDM parameter space inferred from small-scale structure observations.
To maintain consistency, the scenario discussed here would require $g_{\nu\phi}$ to be at least three orders of magnitude below the upper limit shown in Fig.~\ref{fig:constraints_on_gnuchi}, which would correspondingly shift the red dashed curves derived from the cosmological constraints on DM–$\nu$ interactions (Eq.~(\ref{eq:sig_nu_chi})) upward, or invalidating the parameter space of SIDM. Another scenario with a light exotic fermonic DM is considered in Ref.~\cite{Agostini:2020cpz}.

In summary, in this work we consider a well-motivated massive real scalar particle, $\phi$, which mediates interactions both among neutrinos and among DM particles.
Using the PandaX-4T data and the complete analysis machinery, we perform the first direct spectral search for the neutrino-scalar coupling constant, $g_{\nu\phi}$, through the double $\beta$-decay with scalar emission, setting the most stringent laboratory constraints for $m_{\phi}$ in the range of 0.8 to 2 MeV.
These results are in conflict with models invoking such interactions to alleviate the Hubble Tension. For $m_{\phi}$ below 1.3 MeV, our results provide complementary constraints on neutrino-scalar interactions compared to those derived from Big Bang Nucleosynthesis.
If the same scalar $\phi$ mediates DM self-interactions, an assumed coupling $g_{\nu\phi}$ can be combined with neutrino-DM scattering cross-section limits from CMB observations to constrain DM-scalar interactions, enabling a rigorous comparison with the self-interacting dark matter parameter space inferred from small-scale structures.

Our study highlights the potential of double $\beta$-decay data to probe the early Universe, the fundamental properties of neutrinos and dark matter, and, more broadly, physics beyond the Standard Model. As a versatile and ultra-sensitive observatory, PandaX will continue to push these boundaries and offer opportunities to a wide range of fundamental questions.

Note added: As we were finalizing this manuscript, we became aware of an independent theoretical work on this same topic(see Ref.~\cite{boudjema:2025}).

This project is supported in part by grants from National Key R\&D Program of China (Nos. 2023YFA1606200, 2023YFA1606201, 2023YFA1606202), National Science Foundation of China (Nos. 12090060, 12090061, 12090062, U23B2070), and by Office of Science and Technology, Shanghai Municipal Government (grant Nos. 21TQ1400218, 22JC1410100, 23JC1410200, ZJ2023-ZD-003, 25ZR1402223). We thank for the support by the Fundamental Research Funds for the Central Universities. We also thank the sponsorship from the Chinese Academy of Sciences Center for Excellence in Particle Physics (CCEPP), Thomas and Linda Lau Family Foundation, New Cornerstone Science Foundation, Tencent Foundation in China, and Yangyang Development Fund. Finally, we thank the CJPL administration and the Yalong River Hydropower Development Company Ltd. for indispensable logistical support and other help. 
This project is also supported in part by the National Science and Technology Council, the Ministry of Education (Higher Education Sprout Project NTU-114L104022-1), and the National Center for Theoretical Sciences of Taiwan, and the Vietnam National Foundation for Science and Technology Development (NAFOSTED) under grant number 103.01-2023.50 (VQT). We thank Frank Deppisch for insightful discussions.

\bibliographystyle{apsrev4-1}
\bibliography{references}

\onecolumngrid
\allowdisplaybreaks
\newpage

\begin{center}{\large \textbf{Supplementary material}} \end{center}

\section{The phase space factor}

\label{Appexdix phase space factor}
The factor ${\cal G}_{0\nu,\phi}$ is defined as 
\be
    \label{eq:phasefactor}
    {\cal G}_{0\nu,\phi} = \frac{a_{0\nu}} {(m_e R_{0})^2} \int  F_0(Z_f, p_1) F_0(Z_f, p_2) d\Omega_{\nu\phi}. 
\ee
Here the normalization factor $a_{0\nu}$ is given by:
\be
    a_{0\nu} = \frac{(G_{F} \cos \theta_{C} g_{A})^4 m_{e}^9}{64 \pi^{5}},
\ee
where $G_{F}$ is the Fermi constant, $\theta_{C}$ is the Cabibbo angle, $g_{A}$ is the axial vector coupling and $m_{e}$ is the electron mass.
The Fermi function accounts for the Coulomb interaction between the outgoing electrons and the final state nucleus of charge $Z_f = Z + 2$, while $p_{i}$ are the momenta for the two outgoing electrons.
An approximate analytic expression for $F_0(Z_f,p)$ can be found in Ref.~\cite{Doi:1985dx}. 
The nuclear radius is parametrized as $R_{0} \simeq 1.2 A^{1/3}$ fm, where $A$ denotes the atomic mass number of the decaying isotope.

The phase-space element is given by
\begin{equation}
    \begin{aligned}
    \label{eq:phase_space_element}
        d\Omega_{\nu\phi} 
        = \frac{\epsilon_1 \epsilon_2 p_1 p_2 }{(2\pi)^4 m_e^{7}} (T_{\beta\beta}^{2} - m_{\phi}^{2})^{1/2}  d\epsilon_1 d\epsilon_2 d\cos\theta_{p_{1}p_{2}},
    \end{aligned}
\end{equation}
where $\epsilon_{i}$ denotes the energy of each electron with $T_{\beta\beta}=Q_{\beta\beta} + 2m_{e}-\epsilon_{1} -\epsilon_{2}$ the transition energy and $\theta_{p_{1}p_{2}}$ is the crossing angle between the two electron's momenta.

\section{The nuclear matrix element}
\label{Appendix NME}
As the interaction between the real scalar mediator and neutrinos is given by:
\begin{equation}
    \label{eq:Lagrangian}
    {\cal L} = g_{\nu \phi} \bar{\nu}P_{R}\nu^{c} \phi + {\rm h.c.,}
\end{equation}

the $S$ matrix for this process can be calculated as:
\begin{equation}
    \begin{aligned}
        S \sim & \int d^{4}x \int d^{4}y \int d^{4}z \int  \frac{d^{4} q_{1}}{(2\pi)^{4}} \int \frac{d^{4} q_{2}}{(2\pi)^{4}} \sum_{N} \langle F \vert \mathcal{J}^{\rho\dagger}_{L}(\vec{x}) \vert N \rangle \langle N \vert \mathcal{J}^{\sigma\dagger}_{L}(\vec{y}) \vert I \rangle e^{i\epsilon_{1} x_{0}}e^{i\epsilon_{2} y_{0}} \\
        & \times \frac{e^{-iq_{1}(x-z)}}{q_{1}^{2} - m_{1}^{2}} \langle p_{\phi} \vert \phi(z) \vert 0 \rangle \frac{e^{-iq_{2}(z-y)}}{q_{2}^{2} - m_{2}^{2}} \bar{\psi}(\vec{x},\epsilon_{1}) \gamma_{\rho} (1-\gamma_{5})(\slashed{q}_{1} + m_{1})P_{R}(\slashed{q}_{2} + m_{2})(1-\gamma_{5}) \gamma_{\sigma} \psi^{c}(\vec{y},\epsilon_{2}),
    \end{aligned}
\end{equation}
here $\vec{x}, \vec{y}, \vec{z}$ are position operators. $\langle F | \mathcal{J}^{\rho\dagger}_{L}(\vec{x}) | N \rangle \langle N | \mathcal{J}^{\sigma\dagger}_{L}(\vec{y}) | I \rangle$ is the hadronic current, with $I$, $F$ and $N$ being the initial, final and generic intermediate nuclear state, respectively. It can be computed with the impulse approximation~\cite{Engel:2016xgb}. 
The $q_{1,2}(m_{1,2})$ are 4-momenta(masses) for the exchanged neutrinos, and in our model, we only consider one generation of neutrinos (electron neutrinos), thus $m_{1} = m_{2} = m_{\nu_{e}}$. And $\psi$ is the relativistic Coulomb wave function of electrons with energies of $\epsilon_{1,2}$, while $\langle p_{\phi} \vert \phi(z) \vert 0 \rangle = e^{ip_{\phi}z}$ is the plane-wave solution describing the free propagation of the massive scalar field and $p_{\phi}$ the momentum of the scalar field.

The leptonic algebra inside the S matrix can be decomposed into:
\begin{equation}
    \begin{aligned}
    \label{leptonic algebra}
         (1-\gamma_{5})(\slashed{q}_{1} + m_{1})P_{R}(\slashed{q}_{2} + m_{2})(1-\gamma_{5}) & = [q_{1}^{2} - m_{1}^{2} + q_{2}^{2} - m_{2}^{2}](1-\gamma_{5}) \\
         & + [ ( [\slashed{q}_{1}, \slashed{q}_{2}] + (q_{1} - q_{2})^{2} ) + (m_{1} - m_{2})^{2} + (m_{1} + m_{2})^{2}](1-\gamma_{5}),
    \end{aligned}
\end{equation}
and we denote the contribution to the $S$ matrix from the first line in (\ref{leptonic algebra}) as ${\cal{K}}^{(1)}$ and the second line as ${\cal{K}}^{(2)}$. Ref.~\cite{Doi:1987rx} has a detailed analysis of these terms with a massless scalar case, but as we are in the MeV scale, the conclusion that ${\cal{K}}^{(2)}$ part only contributes subdominatly still holds for our massive case. For simplicity, we only consider the contributions from ${\cal{K}}^{(1)}$.

After performing the integration over $z$ and $q_{2}$, we get:
\begin{equation}
    \begin{aligned}
        S \sim & \int d^{4}x \int d^{4}y \int  \frac{d^{4} q}{(2\pi)^{4}} \frac{e^{-iq(x-y)}}{q^{2} - m^{2}} \sum_{N} \langle F \vert \mathcal{J}^{\rho\dagger}_{L}(\vec{x}) \vert N \rangle \langle N  \vert \mathcal{J}^{\sigma\dagger}_{L}(\vec{y}) \vert I \rangle \cdot \langle p_{\phi} \vert \phi(z) \vert 0 \rangle  \\
        & \times \bar{\psi}(\vec{x},\epsilon_{1}) \gamma_{\rho} (1-\gamma_{5}) \gamma_{\sigma} \psi^{c}(\vec{y},\epsilon_{2}) \frac{1}{2}(e^{i p_{\phi} \cdot x} + e^{i p_{\phi} \cdot y}) e^{i(\epsilon_{1} x_{0} + \epsilon_{2} y_{0})},
    \end{aligned}   
\end{equation}
here we changed the integrated momentum from $q_{1}$ to $q$. This form is essentially the same as the standard neutrinoless double beta decay formula with the only difference given by the factor $e^{i p_{\phi} \cdot x} + e^{i p_{\phi} \cdot y}$\cite{Doi:1985dx}. Then by performing the integration with respect to $x^{0},y^{0}$ and $q^{0}$ , we find the $S$ matrix in the closure approximation as:
\begin{equation}
    \begin{aligned}
    \label{S matrix}
        S \sim & \int d^{3}x \int d^{3}y \int  \frac{d^{3} q}{2(2\pi)^{3} } \frac{e^{i \vec{q}\cdot (\vec{x} - \vec{y})}}{w_{q}} \sum_{c = x,y} e^{-i \vec{p}_{\phi}\cdot\vec{c}} \left( \frac{1}{w_{q} + E_{d}} + \frac{1}{w_{q} + E_{d} + E_{\phi}}\right) \sum_{N} \langle F \vert \mathcal{J}^{\rho\dagger}_{L}(\vec{x}) \vert N \rangle \langle N  \vert \mathcal{J}^{\sigma\dagger}_{L}(\vec{y}) \vert I \rangle \\
        & \times \bar{\psi}(\vec{x},\epsilon_{1}) \gamma_{\rho} (1-\gamma_{5}) \gamma_{\sigma} \psi^{c}(\vec{y},\epsilon_{2}) ,
    \end{aligned}     
\end{equation}

and we define the ${\cal{K}}^{(1)}$ factor as:
\begin{equation}
    \begin{aligned}
        {\cal K}^{(1)} & = \int \frac{d^3q}{(2\pi)^3} e^{i \vec{q}\cdot (\vec{x} - \vec{y})} \frac{1}{w_{q}} \cdot \sum_{c = x,y} e^{-i \vec{p}_{\phi}\cdot\vec{c}} \left( \frac{1}{w_{q} + E_{d}} + \frac{1}{w_{q} + E_{d} + E_{\phi}}\right).
    \end{aligned}
\end{equation}
Here $E_{\phi} = \sqrt{\vec{p}_{\phi}^{2} + m_{\phi}^{2}}$ is the energy for the massive real scalar. The empirical value for the energy denominator $E_{d} = \langle E_N\rangle - \tfrac{1}{2}(E_i + E_f) \simeq 1.12 A^{1/2}$ is the same adopted by Ref.~\cite{Haxton:1984ggj}, with $\langle E_N\rangle$ the average energy of the intermediate nuclei states, $E_{i,f}$ the energy for initial/final nuclei states. 
The notation $w_{q} = \sqrt{|\vec{q}|^{2} + m_{\nu_{e}}^2}$ indicates the energy of an electron neutrino with 3-momentum $\vec{p}$.
Note that $E_{\phi} = \sqrt{|\vec{p}_{\phi} |^{2} + m_{\phi}^2}$ should below the transition energy $Q_{\beta\beta} \lesssim \mathcal{O}(\text{MeV})$ for PandaX-4T experiment to be sensitive and the atomic radius for $^{136}Xe$ is $r_{Xe} = r_{0}A^{1/3}$ with $r_{0}\sim1.1-1.3 \; \text{fm}$. 

Since the conditions $|\vec{p}_{\phi}\cdot\vec{x}| \sim |\vec{p}_{\phi}\cdot\vec{y}| \sim \text{MeV}\cdot \text{fm} \ll 1$ and $E_{\phi}\ll w + E_d$ hold, we can neglect the dependence on $E_{\phi}$ and $\vec{p}_{\phi}$ in the above integral:
$e^{-i\vec{p}_{\phi}\cdot\vec{c}} \sim 1$ and $\frac{1}{w_{q} + E_{d}} + \frac{1}{w_{q} + E_{d} + E_{\phi}} \sim  \frac{2}{w_{q} + E_{d}}$. Under these approximations, the ${\cal K}^{(1)}$ reduces to the same formula as the standard neutrinoless double beta decay matrix element up to a normalization factor:
\begin{equation}
    \begin{aligned}
        {\cal K}^{(1)} & \sim \int \frac{d^3q}{(2\pi)^3} e^{i \vec{q}\cdot (\vec{x} - \vec{y})} \frac{1}{w_{q}(w_{q} + E_{d})} .
    \end{aligned}
\end{equation}

With the constants introduced in SMa.I, making construction between the Gamma matrices, one can rewrite the first line in Eq. (\ref{S matrix}) which is just the same formula of the $0\nu\beta\beta$ NME:
\begin{equation}
    |M_{\nu\phi}| = 
         \frac{2\pi R_0}{g_A^2} \int d^{3}x \int d^{3}y 
         \sum_{N}  {\cal K}^{(1)}(\vec{x}, \vec{y})
          \langle F \vert \mathcal{J}^{\rho\dagger}_{L}(\vec{x}) \vert N \rangle \langle N \vert \mathcal{J}^{\dagger}_{\rho,L}(\vec{y}) \vert I \rangle.
\end{equation}

\section{Neutrino -- DM scattering cross section}
\label{Appexdix nuDM scattering}
In this section, we provide the formula of the elastic scattering cross section between neutrino and fermionic DM for the light mediator scenario. The cross section is given as 
\be
\label{eq:sigmel}
\sigma_{\nu \chi}^{\rm el} = \frac{g_{\nu\phi}^{2} g_{\chi\phi}^{2}}{4 \pi} \left[ \frac{1}{s} + \frac{m_{\phi}^{2} - 4 m_{\chi}^{2}}{ m_{\phi}^{2} s + (s - m_{\chi}^{2})^{2} } + \frac{2 (m_{\phi}^{2} - 2 m_{\chi}^{2})}{(s - m_{\chi}^{2})^{2}} \log \left( \frac{m_{\phi}^{2} s }{m_{\phi}^{2} s + (s-m_{\chi}^{2})^{2}} \right)  \right],
\ee
where $s$ is the squared center-of-mass energy.
To obtain a useful approximation of Eq.~\eqref{eq:sigmel}, we introduce the dimensionless variable
\begin{equation}
y \equiv \frac{s - m_{\chi}^{2}}{m_{\chi}^{2}}
= \frac{2 E_{\chi} E_{\nu} (1 - \beta \mu)}{m_{\chi}^{2}},
\end{equation}
where $E_{\chi}$ and $E_{\nu}$ are the energies of the DM and the neutrino, respectively, $\beta$ is the DM velocity, and $\mu \equiv \cos\theta$, with $\theta$ being the scattering angle in the center-of-mass frame.
In the limit of non-relativistic DM, $\beta \to 0$, and the above expression simplifies to
$y \simeq 2 E_{\nu}/m_{\chi}$.
In terms of the new variable $y$, the cross section in Eq.~\eqref{eq:sigmel} becomes 
\be
\label{eq:sigmel_y}
\sigma_{\nu \chi}^{\rm el} = \frac{g_{\nu\phi}^{2} g_{\chi\phi}^{2}}{4 \pi} 
\left[ \frac{1}{m_\chi^2 (1 + y)} 
+ \frac{m_{\phi}^{2} - 4 m_{\chi}^{2}}{ m_{\chi}^{2} [m_\chi^2 y^2 + m_\phi^2 (1 + y)] } 
+ \frac{2 (m_{\phi}^{2} - 2 m_{\chi}^{2})}{m_\chi^4 y^2} \log \left( \frac{m_{\phi}^{2} (1+ y) }{m_\chi^2 y^2 + m_\phi^2 (1 + y)} \right)  \right],
\ee

For low neutrino energies, corresponding to $y \ll 1$, the elastic scattering cross section in Eq.~\eqref{eq:sigmel_y} can be expanded as
\begin{equation}
\sigma_{\nu \chi}^{\rm el}
\simeq
\frac{g_{\nu\phi}^{2} g_{\chi\phi}^{2} \, m_{\chi}^{2}}{2 \pi m_{\phi}^{4}} \, y^{2}
+ \mathcal{O}(y^{3})
\;\propto\;
E_{\nu}^{2},
\end{equation}
demonstrating the characteristic quadratic dependence of the cross section on the neutrino energy in the low-energy limit. We stress that the expansion above assumes a light but finite mediator, such that the mediator mass regulates the infrared behavior while $y \ll 1$.

\end{document}